\documentclass[reprint,onecolumn,longbibliography,floats,aps,amsmath,amssymb,nofootinbib,prd,notitlepage]{revtex4-1}
\usepackage[a4paper,top=3cm,bottom=3cm,left=2cm,right=2cm,marginparwidth=1.75cm]{geometry}
\usepackage[english]{babel}
\usepackage[utf8]{inputenc}
\usepackage{amsmath}
\usepackage{amssymb}
\usepackage{lmodern}
\usepackage{color}
\usepackage{nccmath}
\usepackage{graphicx}
\usepackage{wrapfig}
\usepackage{enumerate}
\usepackage{geometry}
\usepackage{hyperref}
\usepackage{braket}
\usepackage{float}
\usepackage{subfig}
\usepackage{dsfont}

\definecolor{darkmagenta}{rgb}{0.55, 0.0, 0.55}

\definecolor{darkolivegreen}{rgb}{0.33, 0.42, 0.18}

\newcommand{\Weyl}[1]{{\boldsymbol{:}\! #1 \!\boldsymbol{:}}}
\newcommand{\EG}{{\rm EG}}
\newcommand{\PBG}{{\rm PB}^G}

\begin{document}

\title{Revisiting semiclassical effective dynamics for quantum cosmology}
\author{Maciej Kowalczyk}
	\email{maciej.kowalczyk@uwr.edu.pl}
	\affiliation{University of Wroc{\l}aw, Faculty of Physics and Astronomy, Institute for Theoretical Physics, pl. M. Borna 9, 50-204  Wroc{\l}aw, Poland}
\author{Tomasz Paw{\l}owski}
	\email{tomasz.pawlowski@uwr.edu.pl}
	\affiliation{University of Wroc{\l}aw, Faculty of Physics and Astronomy, Institute for Theoretical Physics, pl. M. Borna 9, 50-204  Wroc{\l}aw, Poland}
\begin{abstract}
  We revise the technique of semiclassical effective dynamics, in particular reexamining the evaluation of Poisson structure of the so-called central moments capturing quantum corrections, providing a systematic, pedagogical, and efficient algorithm for evaluation of said structure. The resulting closed formulae for Poisson brackets involve less summations than recent results in the literature, thus being more optimal for applications. Found formulae are then applied to a general class of isotropic cosmological models with locally observable configuration variables for the admitted matter fields. In particular, this allowed to formulate a consistent and nontrivial limit or fiducial cell (infrared regulator) removal for models describing spatially noncompact spacetimes. 
\end{abstract}

\maketitle

\section{Introduction}
\label{sec:intro}

The theory of cosmological perturbations \cite{Dodelson:2003ft,*Mukhanov:2005sc} is one of the best tools to date in trying to understand the physics of the early Universe and its effect on present observable Cosmos. Although in its foundation the theory is classical, the fluctuations it describes are expected (and assumed) to be of quantum origin. Their structure is in turn studied by the techniques of quantum field theory (QFT) on curved spacetime. Although the latter encompasses the quantumness of matter, it is believed that a consistent framework describing the origin of the Universe needs to account for the quantumness of gravity (or the spacetime) itself. There are more than a few attempts at constructing such framework, Geometrodynamics \cite{DeWitt:1967yk}, String Theory \cite{Becker:2006dvp} or Loop Quantum Gravity (LQG) \cite{Thiemann:2007pyv} being among the most prominent. Their simplification to cosmological scenarios allowed to complete their respective quantization programs and extract physical answers (see for example \cite{Ashtekar:2011ni} for a simplification of LQG known as loop quantum cosmology - LQC). At least some of these frameworks have permitted to include perturbations (see again an example \cite{Agullo:2012sh} from LQC) though reconciling them with standard QFT based perturbative formalism requires vast simplifications often restricting the effects of quantum gravity to leading order behavior. Building a more precise interface between these two descriptions requires a formalism which, while classical in nature, allows one to account for quantum (geometry) corrections up to arbitrary order.

One particularly promising framework originates from the geometric formulation of quantum mechanics \cite{Kibble:1978tm,MR1194023,Ashtekar1999,Brody:1999cw,Heydari:2015yqa}. In contrast with traditional quantum mechanics, which is founded on operator algebras and linear vector spaces, with a strong analytical focus rooted in functional analysis, the geometric picture views the Hilbert space as a symplectic manifold endowed with a metric and a naturally emerging Poisson structure, aligning it more closely with the geometric methods used in classical Hamiltonian mechanics. A method well-suited for analyzing quantum dynamics within this framework involves the so-called \emph{Hamburger decomposition}. This technique encodes the semiclassical state into an infinite set of the so-called \emph{central moments} \cite{Bojowald:2005cw,Bojowald:2010qm}, which are real-valued functions of the expectation values of appropriately chosen observables (the so-called \emph{absolute moments}). The dynamics of these moments is then described by a corresponding set of effective equations of motion which take the form of the Hamilton equations for a classical system. This method naturally includes quantum corrections to arbitrary order, allowing for the practical application of a well-defined cutoff to reduce the otherwise infinite (countable) set of equations to a manageable finite one.

In quantum cosmology, this method has already been applied, with a certain level of success \cite{Bojowald:2005cw,Bojowald:2010qm,Brizuela:2021byi}, although its use for more interesting scenarios poses a serious technical challenge. In particular, when dealing with more than one pair of observables, calculating the Poisson bracket between two central moments may become a tedious task prone to errors. To reduce the risk of errors, a more robust approach is to perform the calculation algorithmically, although this method carries an extensive computational cost, often significantly higher than the integration of the resulting equations of motion. As it increases exponentially with the number of classical degrees of freedom of the system, it becomes unpractical for more interesting applications (like perturbative cosmology). 

The above problem has been tackled from an analytical perspective. In particular for central moments 
built over a set of commuting canonical pairs (of observables) forming a Heisenberg algebra the generating function techniques have been applied, see \cite{Bojowald:2005cw,Bojowald:2010qm} and more recent works in the context of cosmological perturbations in \cite{Ding:2023zfk}. Furthermore, a closed-form expression for the Poisson brackets has been obtained through calculations in the algebra
of absolute moments, as presented in \cite{Brizuela:2023ris}. Here we reexamine generating function approach, performing step-by-step derivation of the Poisson structure of central moments built over a set of commuting canonical pairs (of observables) forming the same Heisenberg algebra. The purpose of this work is $(i)$ finding closed formulas for the Poisson structure of the central moments algebra more suitable for the task of building the equations of motion for semiclassical systems (that is minimizing the number of summations involved in evaluating the Poisson brackets) and $(ii)$ providing a systematic and pedagogical framework suitable for further generalization to semiclassical descriptions built over sets of observables forming algebras different than the Heisenberg one.

The paper is organized as follows. First, in sec.~\ref{sec:CentrMoments} we provide a systematic pedagogical discussion on reexamining and rederiving the Poisson brackets, of which an actual step-by-step procedure is performed in Appendix.~\ref{appendixA}. Next, in sec.~\ref{sec:CosmLimit} we apply the found Poisson structure to the case of a general isotropic Friedmann-Lemaître-Robertson-Walker (FLRW) geometry quantized by means of geometrodynamics, admitting a set of matter fields satisfying a very mild assumption, that the configuration variables preset in action are locally observable quantities. In particular, we study the so-called \emph{fiducial cell removal limit}: for models describing homogeneous spacetimes with noncompact spatial homogeneity slices (like flat FRLW Universe) in order to define a Hamiltonian and the momenta, one introduces a certain kind of an infrared regulator -- a finite cell constant in comoving coordinates. Upon quantization, certain aspect of the description become dependent on that choice and the correct theory should only be recovered in the limit, when that cell encompasses the entire universe. We use the semiclassical framework to study that limit, in particular identifying the semiclassical description within that limit, showing in particular that it remains consistent and nontrivial. Finally, in Section \ref{sec:Concl} we present our conclusions and outline directions for future research.

\section{Semiclassical effective dynamics}\label{sec:CentrMoments}

To start with, let us recall the construction of the semiclassical effective dynamics. For a more detailed description we refer the reader to for example \cite{Bojowald:2005cw}.
Consider an arbitrary Hilbert space on which we introduce a set of $N$ pairs of fundamental operators: ``configurations`` denoted as $\hat{X}_j$ and their conjugated momenta  $\hat{Y}_i$ which will form (within each pair) a Heisenberg algebra (while commuting between pairs) $[\hat{X}_j, \hat{Y}_i]=iA_j\delta_{i,j}$. Here we introduce a general coefficient $A_j$ as we may incorporate the dependence of Planck constant $\hbar$ into one of the listed operators in order to permit using this description in more exotic cases.
 
While we start within the framework of quantum theory, it is essential to recognize that every Hilbert space equipped with an inner product naturally acquires the structure of a  K{\"a}hler space \cite{Ashtekar1999,Heydari:2015yqa}. This geometric perspective endows the Hilbert space with a symplectic structure, making it possible to view the Hilbert space as a manifold with an inherent symplectic geometry. In this formulation, quantum states are interpreted as points on this classical phase space but for physical significance in quantum theory are only classes of equivalence of states with respect to relation 
\begin{equation}
  \ket{\Psi_1}\sim\ket{\Psi_2} \Leftrightarrow \ket{\Psi_1}=\lambda\ket{\Psi_2},\qquad \lambda\in \mathbb{C}\setminus \{0\}.
\end{equation}
It is exactly the space of these classes (called \emph{rays}) on which we try to build a coordinate system well-defined on a sufficiently large portion of the physically interesting states. In the first step we restrict our consideration to the states localized in chosen fundamental observables $\hat{X}_i,\hat{Y}_i$, that is their expectation $\braket{\hat{X_j}}, \braket{\hat{Y_j}}$ values are finite. 

Thanks to that restriction these expectation value $X_j:=\braket{\hat{X_j}}, Y_j:=\braket{\hat{Y_j}}$ now form a set of coordinates on the quantum phase space, though such a system is not complete - one cannot reproduce the state just by knowing $X_j$ and $Y_j$. One of the possible ways of completing the coordinate system originates from the \emph{Hamburger decomposition} and is based on the construction of the Central moments \cite{Bojowald:2010qm,Bojowald:2005cw} defined per analogy with such of statistical mechanics, namely
\begin{equation}\label{eq:central-moments}\begin{split}
G^{a_1,..,a_N}_{b_1,...,b_N} &:=\sum_{\substack{k_1=0,..,k_N=0\\ m_1=0,..m_N=0}}^{\substack{a_1,...,a_N\\b_1,...,b_N}}\prod_{j=1}^NB_{k_j,m_j}^{a_j,b_j}X_j^{a_j-k_j}Y_j^{b_j-m_j}\braket{\hat{X_1}^{k_1}\hat{Y_1}^{m_1}...\hat{X_N}^{k_N}\hat{Y_N}^{m_N}}_{\rm Weyl} \\
    &\hphantom{:}={}" \braket{(\hat{X_1}-X_1)^{a_1}(\hat{Y_1}-Y_1)^{b_1}...(\hat{X_N}-X_N)^{a_N}(\hat{Y_N}-Y_N)^{b_N}} {}" \ , \\
    B_{k_j,m_j}^{a_j,b_j} &:=(-1)^{a_j+b_j-k_j-m_j}\binom{a_j}{k_j}\binom{b_j}{m_j}
\end{split}\end{equation}
where the second line is the intuition of the definition, out of which a binomial decomposition infers the precise one (the righthand side in the first line) and $B_{k_j,m_j}^{a_j,b_j}$ are binomial coefficients and by subscript Weyl (further denoted by $\Weyl{\cdot}$) we shall understand total symmetry ordering of the operators given by the McCoy formula
\begin{equation}
  :\hat{X}_j^{n_j}\hat{Y}_j^{k_j}:=\frac{1}{2^{n_j}}\sum_{i=0}^{n_j}\binom{n_j}{i}\hat{X}_j^i\hat{Y}_j^{k_j}\hat{X}_j^{n_{j}-i}=\frac{1}{2^{k_j}}\sum_{i=0}^{k_j}\binom{k_j}{i}\hat{Y}_j^i\hat{X}_j^{n_j}\hat{Y}_j^{{k_j}-i},
\end{equation}
where we utilize the commutativity of operators from different pairs.

In order for the above set to be well defined, the expectation values present there (further called \emph{absolute moments}
\begin{equation}\label{eq:gen-F-def}
  F^{a_1,..,a_N}_{b_1,...,b_N} 
  := \braket{\hat{F}^{a_1,..,a_N}_{b_1,...,b_N}}
  := \braket{\hat{X_1}^{a_1}\hat{Y_1}^{b_1}\ldots\hat{X_N}^{a_N}\hat{Y_N}^{b_N}}_{\rm Weyl} 
  = \braket{(\hat{F}^{a_1,b_1})_1\ldots(\hat{F}^{a_N,b_N})_N} \ , 
\end{equation}
where
\begin{equation}\label{eq:Fj-def}
  (\hat{F}^{a,b})_j\,:=\,\,:\!\hat{X_j}^{a}\hat{Y_j}^{b}\!: \ ,
\end{equation}
must remain finite, which further restricts the domain covered by constructed coordinate system. We strengthen this restriction even further, requiring that the central moments $G^{a_1,..,a_N}_{b_1,...,b_N}$ decrease in magnitude as their order defined as $\sum_{k=1}^N(a_k+b_k)$ increases. These conditions together provide a somewhat strengthened notion of semiclassicality. From now on we will restrict our studies to these states only.

The first step in the program of determining the dynamics of the system is providing the algebra structure of the coordinates $X_j,Y_J,G^{a_1,..,a_N}_{b_1,...,b_N}$. While for $X_j,Y_J$ it is already given by the classical Heisenberg algebra structure, for the remaining quantities it is inherited from the algebra of operators $\hat{F}^{a_1,..,a_N}_{b_1,...,b_N}$. More precisely, it is enough to select a subset of operators $(\hat{F}^{a,b})_j$, for which a Poisson algebraic structure is then uniquely determined by the following condition
\begin{equation}\label{eq:absolute-condition}
  \{\braket{(\hat{F}^{a,b})_j},\braket{(\hat{F}^{c,d}})_j\} = -\frac{i}{\hbar} \braket{[(\hat{F}^{a,b})_j,(\hat{F}^{c,d})_j]}  .
\end{equation}
As general absolute moments are products of $(\hat{F}^{c,d})_j$, that is
\begin{equation}\label{eq:F-dec}
  \hat{F}^{a_1,..,a_N}_{b_1,...,b_N} = (\hat{F}^{a_1,b_1})_1\ldots(\hat{F}^{a_N,b_N})_N \ ,  
\end{equation}
the above bracket determines the whole algebra structure of the (general) absolute moments.

On the technical level, a crucial step in handling the central moments algebra is the derivation of a closed formula for the bracket on the lefthand side of \eqref{eq:absolute-condition} as a function of $(X_j, Y_j, (F^{a,b})_j:=\braket{(\hat{F}^{a,b})_j})$. To achieve this, we first need to compute the commutator between operators $(\hat{F}^{a,b})_j$. The computation can be approached systematically by expanding these operators according to their definitions and utilizing the commutation relations between the fundamental operators.  This approach is particularly effective for evaluating the commutator in two special cases: between any arbitrary  $(\hat{F}^{a,b})_j$ and $(\hat{F}^{1,0})_j=\hat{X}_j$ or $(\hat{F}^{0,1})_j=\hat{Y}_j$. These cases yield the following results
\begin{equation}\label{eq:X-F-comm}\begin{split}
[\hat{X_j},(\hat{F}^{k,n})_j]&= \frac{1}{2^k}\sum_{l=0}^k\binom{k}{l}\hat{X_j}^{k-l}[\hat{X_j},\hat{Y_j}^n]\hat{X_j}^l=iA_jn_j\frac{1}{2^k}\sum_{l=0}^k\binom{k}{l}\hat{X_j}^{k-l}\hat{Y_j}^{n-1}\hat{X_j}^l=iA_jn_j(\hat{F}^{k,n-1})_j\\
[\hat{Y_j},(\hat{F}^{k,n})_j]&= \frac{1}{2^n}\sum_{l=0}^n\binom{n}{l}\hat{Y_j}^{n-l}[\hat{Y_j},\hat{X_j}^k]\hat{Y_j}^l=-iA_jk_j\frac{1}{2^n}\sum_{l=0}^n\binom{n}{l}\hat{Y_j}^{n-l}\hat{X_j}^{k-1}\hat{Y_j}^l=-iA_jk_j(\hat{F}^{k-1,n})_j
\end{split}\end{equation}

It becomes evident that applying this method to compute commutators in the general case quickly becomes a laborious task. To simplify the process and make it more manageable, we employ the concept of generating function \cite{wilf1990generatingfunctionology}. This approach not only simplifies the algebraic manipulations but also provides a more efficient framework for handling the complexities involved. Specifically, in our context, we will use the exponential generating function, defined as follows
\begin{equation}
  \EG_F(\alpha)=\sum_{a,b=0}^{\infty}\frac{(i\alpha_X)^a}{a!}\frac{(i\alpha_Y)^b}{b!} (\hat{F}^{a,b})_j=\exp(i\alpha_X \hat{X}_j + i\alpha_Y\hat{Y}_j)
\end{equation}
To compute the commutator of interest while maintaining the flow of the argument, we outline the procedure as follows and provide a detailed calculation in Appendix \ref{appendixA-F}
\begin{enumerate}
    \item  We begin by considering the commutator of two generating functions. The first step involves expressing each generating function through its series expansion. Consequently, the commutator can be formulated as a summation over four indices, incorporating terms from each generating function along with the commutator of the Absolute Moments. This is given by
    \begin{equation}
  [\EG_F(\alpha),\EG_F(\beta)]=\sum_{a,b,c,d=0}^{\infty}\frac{(i\alpha_X)^a}{a!}\frac{(i\alpha_Y)^b}{b!}\frac{(i\beta_X)^c}{c!}\frac{(i\beta_Y)^d}{d!}[(\hat{F}^{a,b})_j,(\hat{F}^{c,d})_j]
\end{equation}
    \item In the next step, we apply the Baker-Campbell-Hausdorff (BCH) formula, which provides a compact expression for the commutator of exponentiated operators. By applying this formula, we obtain
    \begin{equation}
   [\EG_F(\alpha),\EG_F(\beta)]=-2i\sin(\frac{A_j}{2}(\alpha_X\beta_Y-\alpha_Y\beta_X))\EG_F(\alpha+\beta)
\end{equation}
This result illustrates that the commutator of the generating functions is expressed as a product involving the generating function evaluated at the sum of these parameters scaled by the sine function of the parameters 
$\alpha$ and $\beta$. By performing a Taylor expansion of the latter around zero and subsequently applying the series expansion of the generating functions we arrive to an identity polynomial in coefficients $\alpha,\beta$. Coefficients of this polynomial provide a nontrivial relation between the commutators of particular absolute moments and certain function of these moments.
    \item To identify the coefficients in the above identity, we need to first rearrange the terms in the solution resulting from applying the BCH formula.
    This involves reordering the summation so that the summations over powers of $\alpha$ and $\beta$ are the leftmost one. Upon this reordering, we can extract the terms in front of particular powers of $(\alpha,\beta)$ which yield the identities
    \begin{equation}\label{eq:F-commutator}
       [(\hat{F}^{a,b})_j,(\hat{F}^{c,d})_j]= - i \sum_{r}\frac{A_j^{2r+1}}{4^r}k^{r}_{a,b,c,d} (\hat{F}^{a+c-2r-1,b+d-2r-1})_j
    \end{equation}
with index $r$ satisfying the inequalities 
\begin{equation}\label{eq:r-range} 
  0<r<\frac{1}{2}(\min(a+c,b+d,a+b,c+d)-1)
\end{equation}
(with the demand that the upper limit must be an integer) and coefficients defined as 
\begin{equation}\label{eq:k-def}
  k^{r}_{a,b,c,d}:=\sum_{s=0}^{2r+1}(-1)^s (2r+1-s)!s! \binom{a}{2r+1-s} \binom{b}{s} \binom{c}{s}\binom{d}{2r+1-s}
\end{equation}
\end{enumerate}
By applying \eqref{eq:absolute-condition} to the relation \eqref{eq:F-commutator} we now arrive to a closed formula for the Poisson bracket between two absolute moments
\begin{equation}\label{eq:F-poisson}
    \{\braket{(\hat{F}^{a,b})_j},\braket{(\hat{F}^{c,d}})_j\}=-\frac{A_j}{\hbar}\sum_{r}\frac{A_j^{2r}}{4^r}k^{r}_{a,b,c,d} \braket{(\hat{F}^{a+c-2r-1,b+d-2r-1})_j} \ ,
\end{equation}
with index $r$ range being the same as in \eqref{eq:F-commutator} (in particular satisfying \eqref{eq:r-range}) and coefficients $k^{r}$ given by \eqref{eq:k-def}. Analogously, since for $j\neq k$ $[[(\hat{F}^{a,b})_j,(\hat{F}^{c,d})_k]] = 0$, one can write an analog to \eqref{eq:F-poisson} for general absolute moments $F^{a_1,..,a_N}_{b_1,...,b_N}$ defined in \eqref{eq:gen-F-def}, though we do not present it here due to its complication and limited applicability.

With the algebraic structure of absolute moments now established, we can proceed to probe for an analogous structure of central moments. Given the conceptually straightforward nature and promising results of the previously described procedure, we will adapt and extend it for this goal. In what follows, we outline the general approach and logic, while deferring detailed calculations to an appendix \ref{appendixA-G}.

In constructing the generating function for Central Moments, it is essential to recognize that we are now working with $N$ pairs of operators. Consequently, we must introduce $2N$ parameters to appropriately define the generating function. Per analogy with the generating function for absolute moments we select the following intuition of a definition
\begin{equation}\label{eq:EGG-intuition}
   \EG_G(\alpha)={}" \braket{\exp\Bigg(\sum_{i=1}^N(\alpha_{X_j}(\hat{X}_j-\braket{\hat{X}_j})+\alpha_{Y_J}(\hat{Y}_j-\braket{\hat{Y}_j}))\Bigg)} {}" 
\end{equation}
To realize this intuition we implement two elements: $(i)$ binomial decomposition (analogous to \eqref{eq:central-moments}) in order to precisely express the sum interior, and $(ii)$ the definition of the exponent (of now operator argument) as a Taylor series. Upon that, the generating function takes the form (see Appendix.~\ref{appendixA-G})
\begin{equation}\label{eq:EGG-compact}
    \EG_G(\alpha) = \braket{e^{f_{\alpha}(\hat{X},\hat{Y})}}e^{-\braket{f_{\alpha}(\hat{X},\hat{Y})}} \ , \quad 
    f_{\alpha}(\hat{X},\hat{Y}) := \sum_{j=1}^N(\alpha_{X_j}\hat{X}_j+\alpha_{Y_j}\hat{Y}_J) \ .
\end{equation}
By taking this form of generating function, applying the Leibniz rule to the Poisson bracket and observing that certain components of $\{ \EG_G(\alpha),\EG_G(\beta)\}$ then vanish (see again Appendix.~\ref{appendixA-G}), we obtain the following expression
\begin{equation}\begin{split}
   \{ \EG_G(\alpha),\EG_G(\beta)\} &= \sum_{j=1}^N \frac{A_j}{\hbar}(\beta_{X_j}\alpha_{Y_j}-\alpha_{X_j}\beta_{Y_j})\EG_G(\alpha)\EG_G(\beta)
   \\
   &+\frac{2}{\hbar}\sin\Big(\sum_{j=1}^N \frac{A_j}{2}(\alpha_{X_j}\beta_{Y_j}-\alpha_{Y_j}\beta_{X_j})\Big)\EG_G(\alpha+\beta)
\end{split}\end{equation}
As in the previously derived Poisson bracket for Absolute Moment this expression contains the generating function evaluated at the sum of $\alpha$ and $\beta$ scaled with the more complex arrangement of the sine function. In addition, it contains additional $N$ terms, quadratic in the generating function. By rearranging terms and switching the order of summation we recover the closed formula for the Poisson bracket between two arbitrary Central Moments:
\begin{equation}\begin{split}\label{eq:G-poisson}
&\{G^{a_1,..,a_N}_{b_1,...,b_N}, G^{c_1,..,c_N}_{d_1,...,d_N} \}= \sum_{n=1}^N \frac{A_n}{\hbar}\Big(b_n c_nG^{a_1,..,a_N}_{b_1,...,b_n-1,...,b_N}G^{c_1,..,c_n-1,...,c_N}_{d_1,...,d_N}-a_n d_nG^{a_1,..,a_n-1,...,a_N}_{b_1,...,b_N}G^{c_1,..,c_N}_{d_1,...,d_n-1,...,d_N}\Big)\\
&+\frac{1}{\hbar}\sum_{r}\sum_{s}\frac{(-1)^{r+s}}{4^r}  \sum_{q_1,..,q_N}\delta_{2r+1,q_1+..+q_N}A_j^{q_j} K^{a_i,b_i,c_i,d_i}_{q_i,s} G^{a_1+c_1-q_1,..,a_N+c_N-q_N}_{b_1+d_1-q_1,...,b_N+d_N-q_N} \ .
\end{split}
\end{equation}
The symbol $\delta$ in this context represents the Kronecker delta. While this formula bears a resemblance to the one presented in \cite{Bojowald:2005cw}, it notably differs in the coefficients. Our calculations show they take the following form
\begin{equation}\label{K}
    K^{a_i,b_i,c_i,d_i}_{q_i,s}=\sum_{l_1,...,l_N}\delta_{s,l_1+..+l_N}  \prod^N_{i=1}l_i!(q_i-l_i)!\binom{a_i}{q_i-l_i}\binom{b_i}{l_i}\binom{c_i}{l_i}\binom{d_i}{q_i-l_i} \ ,
\end{equation}
where the coefficients $l_i$ span the set of values satisfying the following constraints:
\begin{subequations}\begin{align}
    &\max(0,q_i-\min(2r+1-s,a_i,d_i))\leq l_i \leq\min(q_i,\min(s,b_i,c_i)) \ , \\
    &0 \leq q_i \leq\min(s,b_i,c_i)+\min(2r+1-s,a_i,d_i) \ , \\
    &0 \leq s \leq\min(2r+1,\sum_{i=1}^N \min(b_i,c_i)) \ , \\
    &0 \leq r \leq \frac{1}{2}\sum_{i=1}^N \min(b_i,c_i)+\frac{1}{2}\sum_{i=1}^N \min(a_i,d_i)-\frac{1}{2} \ ,
\end{align}\end{subequations}
with the requirement that the upper bound for $r$ must be a positive integer.
This form aligns well with recent calculations \cite{Brizuela:2023ris}. 

Finally, to complete the symplectic structure one needs to find the Poisson bracket between 
classical trajectories and Central moments. By implementing relation \eqref{eq:F-poisson} and reordering coefficients in binomial terms it is relatively easy to find that $X_i$ and $ Y_i$ will commute with $G^{a_1,..,a_N}_{b_1,...,b_N}$, that is
\begin{equation}
    \{X_i,G^{a_1,..,a_N}_{b_1,...,b_N}\} = \{Y_i,G^{a_1,..,a_N}_{b_1,...,b_N}\} = 0 .
\end{equation}

Having established a system of coordinates and their Poisson structure, we are ready to describe a prescription for probing the dynamics. This is achieved by recognizing that the expectation value of any sufficiently well-behaved composite observable $\hat{O}:=f(\hat{X_1},\hat{Y_1},..,\hat{X_N},\hat{Y_N})$ can be expressed (via an analog of a Taylor expansion) in terms of central moments.

This applies in particular to the quantum Hamiltonian
\begin{equation}\begin{split}\label{eq:HQ}
  H_Q(X_1,Y_1,..,X_N,Y_N,G^{a_1,...,a_N}_{b_1,...,b_N})& := \braket{H_Q(\hat{X_1},\hat{Y_1},..,\hat{X_N},\hat{Y_N})}_{\rm Weyl}\\ 
  &= \sum_{\substack{a_1=0,..,a_N=0\\b_1=0,...,b_N=0}}^{\infty}\prod_{j=1}^N \frac{1}{a_j!b_j!} \frac{\partial^{a_1+b_1+...+a_N+b_N} H(X_1,Y_1,...,X_N,Y_N)}{\partial X_1^{a_1}  \partial Y_1^{b_1}...\partial X_N^{a_N}\partial Y_N^{b_N} } G^{a_1,..,a_N}_{b_1,...,b_N} 
 \end{split}\end{equation}
 This expression generates a Hamiltonian flow, encapsulating the evolution of states within the quantum phase space. Here, the variables $\{X_1,Y_1,..,X_N,Y_N,G^{a_1,...,a_N}_{b_1,...,b_N}\}$ evolve according to the dynamics induced by a full set of effective equations of motion 
 \begin{equation}\begin{split}
\frac{d}{dt}X_j= \{X_j, H_Q\}, \qquad \frac{d}{dt}Y_j= \{Y_j, H_Q\}, \qquad  \frac{d}{dt}G^{a_1,..,a_N}_{b_1,...,b_N} = \{G^{a_1,..,a_N}_{b_1,...,b_N} , H_Q\}
\end{split}\end{equation}
These equations formally resemble classical Hamiltonian equations of motion; however, they are enriched by the inclusion of terms that encode the back-reaction effects of fundamental quantum operators. In the case of variables forming a standard Heisenberg algebra the contribution of particular moments is scaled with $\hbar^d$, where $d$ is the order of the moments (already defined as the sum of all indices) \cite{Bojowald:2005cw} thus providing a natural hierarchy of quantum corrections. In particular, at zeroth order, we recover the classical system. The first-order terms, due to their very construction identically vanish, thereby preserving the classical structure at this level. Given that equation \eqref{eq:HQ} encompasses an infinite series of variables, it becomes essential to introduce a practical cutoff. The challenge then is to determine the order of this cutoff in an optimal way, that is so that the truncated set of equations will capture the relevant quantum effects without oversimplification while still being manageable.

\section{Application to cosmological models}\label{sec:CosmLimit}

To test the applicability of the formalism introduced in the previous section we wish to apply it to a cosmological scenario. We start by introducing a classical description of the flat, isotropic universe described by the Friedman-Lemaitre-Robertson-Walker metric described by the action:
\begin{equation}\label{eq:action}
    S = \frac{1}{16\pi G} \int_{\mathcal{V}\times I} d^4x \sqrt{-g} \, R -\int_{\mathcal{V}\times I} d^4x \sqrt{-g} \, \mathcal{L}_{\text{m}}
\end{equation}
where the first term is Einstein Hilbert action, $\mathcal{V}$ is the so called \emph{fiducial cell} which
define as the finite region of the Universe that remains constant in co-moving coordinates, and $I$ is some (not yet defined)time interval. To hold generality of our reasoning we introduce an action of an arbitrary matter content (described by $\mathcal{L}_{\text{m}}$) weekly coupled to gravity.
This matter content is most commonly modeled as a barotropic fluid, characterized by a simple linear relationship between energy density $\rho$ and pressure $p$. 
This phenomenological characterization enables us to incorporate a variety of cosmic fluids, from dust to radiation and even dark energy, each influencing the dynamics of the Universe in distinct ways.

At this moment a certain attention needs to be given to the role of $\mathcal{V}$. In case when the homogeneity slices of the Universe are compact, $\mathcal{V}$ simply represents an entire slice. In the case these slices are noncompact, taking them would introduce divergences into \eqref{eq:action}. To avoid that, one restricts the integration over spatial coordinates to a finite spatial region constant in comoving coordinates -- the \emph{finite} fiducial cell. Such cell becomes in certain sense an infrared regulator of the model. Such regularized model may feature a dependence on the choice of $\mathcal{V}$, however the correct physical description should be defined in the \emph{regulator removal limit}, which here corresponds to $\mathcal{V}$ expanding to encompass the whole spatial slice.

From this perspective, it is useful to describe the total energy of the system using the Hamiltonian formalism. In order to do so we shall choose a pair of phase space variables describing a single gravitational degree of freedom, the three-volume of the spatial slice defined as  $V := a^3$ where $a$ is the scale factor and its conjugate momentum, proportional to a Hubble parameter $H=\dot{a}/a$  as $\pi_V=-(4 \pi G )^{-1} H$. Thus the total energy of the system could be described by the following Hamiltonian
\begin{equation}\label{eq:H}
   \mathcal{H} =  -N (6 \pi G V \pi_V^2 +  V \rho) \ ,
\end{equation}
where  $N$ is a lapse function typically set to one, we shall refrain from choosing a specific gauge for now. From the expression of the above Hamiltonian, we can immediately see that it could be decoupled concerning the volume. Moreover, by finding corresponding equations of motions:
\begin{equation}
\dot{V} = -12 \pi G N V \pi_V \ , \qquad \dot{\pi}_V = N \left( 6 \pi G \pi_V^2 - p \right), \qquad \dot{\rho} = 12 \pi G N \pi_V (\rho+p)
\end{equation}
we conclude that our classical system is manifestly scale invariant concerning the three-volume $V$. This should not be a surprise as it is expected that the evolution of the energy density and the Hubble parameter should be independent of the specific volume of the universe. This aligns with the homogeneous and isotropic assumptions, where the dynamics are governed by the Friedman and the continuity equations which in turn could be recreated from the equations of motion given above.

However, if we insist that our starting point is action \eqref{eq:action}, the Lagrangian formulation requires a more detailed, ,,microscopic" description of the degrees of freedom and their interactions in terms of a set of generalized coordinates $q_i$ and their velocities $\dot{q}_i$ to capture the system's dynamics accurately. In cosmological contexts, this is often addressed by using a homogeneous scalar field $\phi$ with arbitrary potential $V(\phi)$. This approach is particularly relevant in inflationary cosmology,  where a kinetic term of the scalar field (often referred to as ``inflaton'') is negligible compared to its potential energy (often modeled by the quadratic \cite{Linde:1983gd} or Starobinsky \cite{Starobinsky:1980te} potential) leading in turn to a quasi-DeSiter evolution of the Universe. For the purposes of our analysis, however, we will refrain from specifying a particular form of the matter content. Furthermore, for the purpose of clarity we simplify by considering a single canonical pair of degrees of freedom, represented by a generalized coordinate $q$ and its conjugated momentum $p_q$. Note though, that our reasoning can be generalized to an arbitrary number of canonical matter degrees of freedom commuting between pairs. Focusing on this approach, we examine a class of systems that, through a rescaling of the momentum as  $\pi_{q} := p_{q}/V$ can be described using a Hamiltonian that preserves scale invariance with respect to the volume
\begin{equation}
    \mathcal{H}=  V\Tilde{H}(\pi_V,q, \pi_{q}) \ ,
\end{equation}
which remains applicable regardless of the specific choice of conventional matter component in FLRW cosmology. Moreover, as we may introduce an additional term to the action \ref{eq:action},  with not any direct physical effects on the system other than providing a time reference, form $\Tilde{H}$ will also depend on the lapse function. In particular, we mention two such choices of external clock irrational dust which corresponds to $N=1$ \cite{Husain:2011tk,Husain:2011tm} and action of the free scalar field $\psi$ for which $N=2V$ \cite{Kowalczyk:2022ajp, Assanioussi:2019iye}.

The Hamiltonian formulation within the context of cosmology serves as a foundational framework for quantization, employing standard quantum mechanical techniques in the Wheeler-DeWitt geometrodynamical approach. In this framework, the Hilbert space associated with the physical system is constructed as a product space of square-integrable functions reflecting wave functions. In this scenario, phase space variables associated with gravitational and matter degrees of freedom are elevated to the status of quantum operators that act on the aforementioned Hilbert space. Specifically, the coordinate system of quantum phase space will be described according to a prescription given in the previous section by the expectation values of each observable, along with an infinite set of central moments
\begin{equation}
    G^{a,c}_{b,d}={}" \braket{(:\hat{V}-V)^{a}(\hat{\pi}_V-\pi_V)^{b}(\hat{q}-q)^{c}(\hat{p}_{q}-p_{q})^{d}:} {}" \ , 
\end{equation}
where $\braket{\hat{V}}:=V $, $\braket{\hat{\pi}_V}:=\pi_V$, $\braket{\hat{q}}:=q$ $\braket{\hat{p}_{q}}:=p_{q}$ are classical trajectories with the evolution generated by the Hamiltonian flow tailored by the expectation value of the quantum Hamiltonian: 
\begin{equation}\label{eq:HQcosmology}
    H_Q= \sum_{\substack{a=0,c=0\\b=0,d=0}}^{\infty}\frac{\partial_V^a\partial^b_{\pi_v}\partial_{q}^c\partial_{p_{q}}^d H(V, \pi_V, q, p_{q})}{a!b!c!d!} G^{a,c}_{b,d} \ .
\end{equation}
Since $H(V, \pi_V, q, p_{q})$ entering above quantum Hamiltonian is a classical function we can once again choose to work with densitized conjugated momentum for the scalar field. However this comes at the cost of deviating the Poisson structure reflected in the modified derivatives:  $\partial_V=\partial_V-\frac{\pi_{q}}{V}\partial_{\pi_{q}}$ and $\partial_{p_{q}}=\frac{1}{V}\partial_{\pi_{q}}$. This re-scaling also impacts the central moments and quantum Hamiltonian 
\begin{equation}
    Q^{a,c}_{b,d}=\frac{1}{V^{a+d}} G^{a,c}_{b,d} \ , 
    \qquad 
    H_Q = \sum_{\substack{a=0,c=0\\b=0,d=0}}^{\infty}\frac{V^{a+d}Q^{a,c}_{b,d} }{a!b!c!d!} (\frac{1}{V}\partial_{\pi_{q}})^d(\partial_V-\frac{\pi_{q}}{V}\partial_{\pi_{q}})^a\partial^b_{\pi_v}\partial_{q}^c \Big(V \Tilde{H}(\pi_V, q, \pi_{q})\Big) \ .
\end{equation}
Although the equation for quantum Hamiltonian may initially appear more complex in compression to \ref{eq:HQcosmology}, leveraging the linear dependence of the decoupled classical Hamiltonian on the volume significantly simplifies the analysis. Specifically, the operator  $\partial_V-\frac{\pi_{q}}{V}\partial_{\pi_{q}}$ can be decomposed as a composite operator 
$AB$ where $A=\frac{1}{V}$ and $B=V\partial_V-\pi_{q}\partial_{\pi_{q}}$. This decomposition enables us to use the commutation relation
\begin{equation}
    [A,B]=A
\end{equation}
in finding arbitrary power of the product $AB$. Applying this commutation relation recursively, we arrive at the expression where all $A$ has been shifted to a left-hand side
\begin{equation}\label{AB}
    (AB)^n=A^a\sum_i k_{ai}B^i \ ,
\end{equation}
where we do not specify the exact form of numeric coefficients $k_{ai}$. Since $B$ acts on functions linear in volume, we can further simplify it as
\begin{equation}
    B=\mathbf{1}-\pi_{\phi}\partial_{\pi_{\phi}} \ ,
\end{equation}
where $\mathbf{1}$ stands for identity operator acting on functions to the right. results in a Quantum Hamiltonian that is explicitly linear in volume:
\begin{equation}
    H_Q= V\sum^{\infty}_{a,c,b,d=0} \Tilde{H}^{a,c}_{b,d}(\pi_V,q,\pi_{q}) Q^{a,c}_{b,d} \ ,
\end{equation}
where we introduce function $\Tilde{H}^{a,c}_{b,d}(\pi_V,q,\pi_{q}) $  depending on coefficients $a,b,c,d$ and observables $\pi_V,q\pi_{q} $.  
Hence in order to obtain quantum trajectories we need to solve a set of coupled equations of motion
\begin{equation}\label{eq:eqm}\begin{split}
   &\dot{V}= V\sum_{\substack{a,c=0\\b,d=0}}^{\infty}\partial_{\pi_V}\Tilde{H}^{a,c}_{b,d}(\pi_V,q,\pi_{q}) Q^{a,c}_{b,d} \ , 
   \qquad 
   \dot{\pi}_V=\sum_{\substack{a,c=0\\b,d=0}}^{\infty} (a+d-1)\Tilde{H}^{a,c}_{b,d}(\pi_V,q,\pi_{q})Q^{a,c}_{b,d} \ ,\\
   &\dot{q}=\sum_{\substack{a,c=0\\b,d=0}}^{\infty}\partial_{\pi_{\phi}}\Tilde{H}^{a,c}_{b,d}(\pi_V,q,\pi_{q}) Q^{a,c}_{b,d} \ , 
   \qquad
   \dot{\pi}_{q}=-\sum_{\substack{a,c=0\\b,d=0}}^{\infty}\partial_{\phi}\Tilde{H}^{a,c}_{b,d}(\pi_V,q,\pi_{q}) Q^{a,c}_{b,d} \ ,\\
   &\dot{Q}^{a,c}_{b,d}=-(a+d)\frac{\dot{V}}{V} Q^{a,c}_{b,d} + V\sum^{\infty}_{i,k,j,m=0} \Tilde{H}^{a,c}_{b,d}(\pi_V,q,\pi_{q}) \{Q^{a,c}_{b,d},Q^{i,k}_{j,m}\} \ ,
\end{split}\end{equation}
where volume enters only two of them. Notably, at zeroth order truncation, the resulting equation of motion reproduces the classical dynamics. At higher orders, particularly for orders larger/ equal to two, quantum effects introduce interactions between geometric and matter observables, potentially revealing back-reaction effects. Since the central moments are constructed using semiclassical states, these quantum back-reaction effects will be most visible at lower truncation orders and very fast-suppressedas the order of truncation increases.

To further clarify the prescription, we provide a more concise formulation for the equation of motion of $Q$. By substituting the equation for $\dot{V}$ and computing Poisson brackets for the subsequent term using the relations established in \eqref{eq:G-poisson}, we derive:
\begin{equation}\label{eq:eqmQ}\begin{split}
    \dot{Q}^{a,c}_{b,d}&=-(a+d)\sum^{\infty}_{i,k,j,m=0}\partial_{\pi_V} \Tilde{H}^{i,k}_{j,m}(\pi_V,q,\pi_{q}) Q^{i,k}_{j,m} Q^{a,c}_{b,d}\\
    &+ \sum^{\infty}_{i,k,j,m=0} \Tilde{H}^{i,k}_{j,m}(\pi_V,q,\pi_{q})\Big(b i Q^{a,b-1}_{c,d} Q^{i-1,j}_{k,m}-a j Q^{a-1,b}_{c,d} Q^{i,j-1}_{k,m}+d k Q^{a,b}_{c,d-1} Q^{i,j}_{k-1,m}-c m Q^{a,b}_{c-1,d} Q^{i,j}_{k,m-1}\Big)\\
    &+ \sum^{\infty}_{i,k,j,m=0} \Tilde{H}^{i,k}_{j,m}(\pi_V,q,\pi_{q})\sum_{r}\sum_{s}\frac{(-1)^{r+s}}{4^r}(\frac{\hbar}{V})^{2r}  \sum_{q_1,
    ,q_2}\delta_{2r+1,q_1+q_2}K^{a,b,i,j}_{q_1,s}K^{c,d,k,m}_{q_1,s} Q^{a+i-q_1,c+k-q_2}_{b+j-q_1,d+m-q_2}
    \end{split}
\end{equation}
and it becomes apparent, that with the above equation replacing the last one in \eqref{eq:eqm}, the latter is asymptotically scale invariant. Specifically, the volume appears only in the equation of motion for itself, while for the remaining variables, it only affects the evolution of moments through factors containing its negative powers. This indicates that the system is well-defined in the limit of the fiducial cell expanding to encompass the whole universe, which corresponds to $V\to\infty$ -- our infrared regulator removal limit. Since these factors are always of the form $(\hbar/V)^{2r}$, this limit is also equivalent to $\hbar\to 0$. thus is equivalent to taking a classical limit of the model, where instead of a quantum state we deal with a statistical distribution of classical ones.

\section{Conclusions}\label{sec:Concl}

In this paper, we revised the recent progress on one of the techniques of determining the dynamics of quantum mechanical systems -- the semiclassical effective mechanics. In particular, for the system featuring a set of fundamental observables as canonical pairs forming Heisenberg algebra, we explored the construction and structure of the central moments and performed detailed calculations of its Poisson bracket. Given the technical challenges of applying this method in scenarios involving multiple observables, we provided a step-by-step derivation of the relevant Poisson brackets by using the generating function technique. The resulting closed formuli are simpler in comparison with those already present in the literature, thus are more promising in applications to the systems involving a larger number of degrees of freedom.

Found formulae for Poisson brackets were subsequently applied to systems describing an isotropic Friedmann-Lemaître-Robertson-Walker (FLRW) cosmology (quantized within the Wheeler-DeWitt geometrodynamical framework), for which general matter content was chosen. The sole requirement of the admitted matter is that the configuration variables for the fields are locally observable quantities (thus making their canonical momenta extensive quantities). For the models describing Universes of noncompact spatial homogeneity slices, we investigated the so-called fiducial cell (or infrared regulator) removal limit. It was shown, that taking this limit provides a nontrivial and consistent description. Furthermore, this limit is consistent with taking the limit $\hbar\to 0$, thus enforces a classicalization of the system. This result generalizes that of \cite{Brizuela:2021byi} established for the inflaton field.

The presented and improved framework not only provides a deeper understanding of the semiclassical dynamics in cosmological settings but also lays the groundwork for extensions to more complex systems, including those involving observables that form non-Heisenberg algebras (for example the algebra of fundamental operators characteristic for loop quantum cosmology and other approaches based on polymeric quantization). Through this work, we aim to enhance the accessibility and applicability of the geometric approach to quantum mechanics, fostering its use in addressing foundational problems in quantum cosmology and beyond.

\begin{acknowledgments}
    This work was supported in part by the Polish National Center for Science (Narodowe Centrum Nauki -- NCN) grant OPUS 2020/37/B/ST2/03604.
\end{acknowledgments}

\appendix

\section{Algebra structure - detailed investigation}\label{appendixA}

Here we present more technically involved steps of evaluating the Poisson brackets between the components of semiclassical description: absolute and central moments.

\subsection{Absolute moments}\label{appendixA-F}

Since the observables corresponding to selected variables commute between distinct canonical pairs in probing the structure of the absolute moments it is enough to consider ``single pair'' ones, that is $(F^{a,b})_j$ defined in \eqref{eq:Fj-def}. As their Poisson structure is defined directly by the commutation relations of the operators through \eqref{eq:absolute-condition}, it is sensible to focus on the latter first. We do so by using the generating function technique, a powerful tool that systematically encapsulates the contributions of each individual moment. Specifically, we construct the exponential generating function $\exp(i\alpha_X \hat{X}_j + i\alpha_Y\hat{Y}_j)$ by assembling it through a series analogous to the Taylor expansion of an exponent
\begin{equation}\label{eq:F-gen}\begin{split}
  \EG_F(\alpha)
  &= \sum_{a,b=0}^{\infty}\frac{(i\alpha_X)^a}{a!}\frac{(i\alpha_Y)^b}{b!} (\hat{F}^{a,b})_j
  = \sum_{a,b=0}^{\infty}\frac{(i\alpha_X)^a}{a!}\frac{(i\alpha_Y)^b}{b!}\frac{1}{2^a}\sum_{n=0}^a\binom{a}{n}\hat{X}_j^n\hat{Y}_j^b\hat{X}_j^{a-n}\\
  &= \sum_{n,b,a=0}^{\infty}\frac{(i\alpha_X)^{n}}{n!}\frac{(i\alpha_X)^{a}}{a!}\frac{(i\alpha_Y)^b}{b!}\frac{1}{2^{a+n}}\hat{X}_j^n\hat{Y}_j^b\hat{X}_j^{a}
  = \exp(\frac{i\alpha_X}{2}\hat{X}_j)\exp(i\alpha_Y\hat{Y}_j)\exp(\frac{i\alpha_X}{2}\hat{X}_j)\\
  &= \exp(i\alpha_X \hat{X}_j + i\alpha_Y\hat{Y}_j)
\end{split}\end{equation}
where we employed the McCoy formula for the totally symmetric ordering of operators, used the definition of the exponent of an operator as Taylor expansion series and finally applied the Baker-Campbell-Hausdorff formula for operators satisfying $[\hat{X},[\hat{X},\hat{Y}]]=0$.

Once a convenient generating function is defined, we can proceed to compute the commutator between two such generating functions. This is approached from two directions. First, we use the series expansion from \eqref{eq:F-gen}:
\begin{equation}\label{eq:EGF-comm}\begin{split}
  [\EG_F(\alpha),\EG_F(\beta)]
  = \sum_{a,b,c,d=0}^{\infty}\frac{(i\alpha_X)^a}{a!}\frac{(i\alpha_Y)^b}{b!}\frac{(i\beta_X)^c}{c!}\frac{(i\beta_Y)^d}{d!}[(\hat{F}^{a,b})_j,(\hat{F}^{c,d})_j]
\end{split}\end{equation}
Second, we apply the Baker-Campbell-Hausdorff (BCH) formula directly to the exponential form of the generating function:
\begin{equation}\label{EGF-comm}\begin{split}
   [\EG_F(\alpha),\EG_F(\beta)]
   &= [\exp(i\alpha_X \hat{X}_j + i\alpha_Y \hat{Y}_j),\exp(i\beta_Y\hat{X}_j + i\beta_Y \hat{Y}_j)]
   \\
   &= -2i\sin(\frac{A_j}{2}(\alpha_X\beta_Y-\alpha_Y\beta_X))\EG_F(\alpha+\beta)
\end{split}\end{equation}
To be able to compare the above equation with \eqref{eq:EGF-comm} we expand its rightmost formula via a Taylor expansion and the binomial theorem:
\begin{equation}\label{eq:EFG-expansion}\begin{split}
  &[\EG_F(\alpha),\EG_F(\beta)]=-2i\sum_{r=0}^{\infty}\frac{(-1)^r}{(2r+1)!}(\frac{A_j}{2})^{2r+1}(\alpha_X\beta_Y-\alpha_Y\beta_X)^{2r+1}\sum_{m,n=0}^{\infty}i^{m+n}\frac{(\alpha_X+\beta_X)^m}{m!}\frac{(\alpha_Y+\beta_Y)^n}{n!}(\hat{F}^{m,n})_j\\
  &=-2i\sum_{r=0}^{\infty}\sum_{s=0}^{2r+1}\frac{(-1)^{r+s}}{(2r+1-s)!s!}(\frac{A_j}{2})^{2r+1}(\alpha_X\beta_Y)^{2r+1-s}(\alpha_Y\beta_X)^s\sum_{a,b,c,d=0}^{\infty}\frac{(i\alpha_X)^a}{a!}\frac{(i\alpha_Y)^b}{b!}\frac{(i\beta_X)^c}{c!}\frac{(i\beta_Y)^d}{d!}(\hat{F}^{a+c,b+d})_j\\
  &=2i\sum_{r=0}^{\infty}\sum_{s=0}^{2r+1}\frac{(-1)^{r+s}}{(2r+1-s)!s!}(\frac{A_j}{2})^{2r+1}\sum_{\substack{a,d=2r+1-s\\c,b=s}}^{\infty}\frac{(i\alpha_X)^a}{(a-2r-1+s)!}\frac{(i\alpha_Y)^b}{(b-s)!}\frac{(i\beta_X)^c}{(c-s)!}\frac{(i\beta_Y)^d}{(d-2r-1+s)!}(\hat{F}^{a+c-2r-1,b+d-2r-1})_j \\
  &= 2\sum_{a,b,c,d=0}^{\infty}\frac{(i\alpha_X)^a}{a!}\frac{(i\alpha_Y)^b}{b!}\frac{(i\beta_X)^c}{c!}\frac{(i\beta_Y)^d}{d!} \sum_{r}(\frac{iA_j}{2})^{2r+1}k^{r}_{a,b,c,d} (\hat{F}^{a+c-2r-1,b+d-2r-1})_j \ ,
\end{split}\end{equation}
where to obtain the third line we rename indexes to bring the interior part of the sum to the form resembling the expression  in \ref{eq:EGF-comm}, then (in the fourth line) we change the order of summation while introducing the coefficient:
\begin{equation}
  k^{r}_{a,b,c,d}:=\sum_{s=0}^{2r+1}(-1)^s (2r+1-s)!s! \binom{a}{2r+1-s} \binom{b}{s} \binom{c}{s}\binom{d}{2r+1-s} \ ,
\end{equation}
with (integer) index $r$ running over the set of values satisfying the inequality $0 \leq r \leq \frac{1}{2}(\min(a+c,b+d,a+b,c+d)-1)$.
Finally, an order-by-order comparison of the last line of \eqref{eq:EFG-expansion} with \eqref{eq:EGF-comm} allows to determine the desired commutators as functionals of the moments themselves:
\begin{equation}
  [(\hat{F}^{a,b})_j,(\hat{F}^{c,d})_j]= - i \sum_{r}\frac{A_j^{2r+1}}{4^r}k^{r}_{a,b,c,d} (\hat{F}^{a+c-2r-1,b+d-2r-1})_j \ .
\end{equation}

\subsection{Central Moments}\label{appendixA-G}

With the algebraic structure of absolute moments established, we can now move on to computing the algebra of central ones. This can be accomplished either by a direct calculation using the known algebra of absolute moments (recently detailed in \cite{Brizuela:2023ris}) or by constructing a generating function for central moments, similarly to the approach used for absolute moments. This latter method, as demonstrated in works such as \cite{Bojowald:2010qm,Bojowald:2005cw} and variation of it in recent \cite{Ding:2023zfk} provides a systematic framework for deriving the desired results. In this paper, we adopt the second approach, presenting a detailed derivation while correcting some minor mistakes identified in \cite{Bojowald:2005cw}.

Let us start with formulating (recalling) the idea behind the generating function technique implemented here. The general intuition is to build a functional of basic observables of which central moments we constructed: $(X_j,Y_J)$ in such a way that its Taylor expansion generates a series whose coefficients are proportional to particular central moments. For absolute moments a particularly convenient choice was the exponent defined in \eqref{eq:F-gen}. Per analogy with that form, here we propose an analog of which intuitive meaning can be captured in the following (at the moment ill-defined) expression 
\begin{equation}\label{eq:EGG-intuition-app}
   \EG_G(\alpha)={}" \braket{\exp\Bigg(\sum_{i=1}^N(\alpha_{X_j}(\hat{X}_j-\braket{\hat{X}_j})+\alpha_{Y_J}(\hat{Y}_j-\braket{\hat{Y}_j}))\Bigg)} {}" \ .
\end{equation}
We begin by giving to this exponent a precise meaning by $(i)$ taking a Taylor expansion of the above exponent (with respect to coefficients $\alpha_{X_j},\alpha_{Y_j}$), $(ii)$ expanding the powers of sums via binomial expansion, and $(iii)$ acting with $\braket{\cdot}$ on the resulting polynomials of $\hat{X}_j,\hat{Y}_j$ (with powers of $\braket{\hat{X}_j},\braket{\hat{Y}_j}$ as multiplicative factors for each term). As a result we obtain the definition of $\EG_G$ in a form of a series involving absolute moments and expectation values $\braket{\hat{X}_j},\braket{\hat{Y}_j}$, giving the desired precise definition
\begin{equation}\label{eq:EGG-predef-app}
   \EG_G(\alpha) := \sum_{\substack{a_1,..,a_N=0\\b_1,..,b_N=0}}^{\infty} \left\langle \prod^N_{j=1}  \frac{(\alpha_{X_j})^{a_j}}{a_j!}\frac{(\alpha_{Y_j})^{b_j}}{b_j!} \sum_{k_j=0}^{a_j}\sum_{l_j=0}^{b_j} (\hat{F}^{k_j,l_j})_j \right\rangle (-1)^{a_j-k_j+b_j-l_j} \braket{\hat{X}_j}^{a_j-k_j} \braket{\hat{Y}_j}^{b_j-l_j} \binom{a_j}{k_j}\binom{b_j}{l_j} \ ,
\end{equation}
where the expectation value is applied to the whole products of $(\hat{F}^{a_j,b_j})_j$ over $j$ for each element of the sums over $k_j,l_j$ separately. As it can be (independently) quite easily inferred from \eqref{eq:EGG-intuition-app} already, the (expectation values of the) absolute moments gather up to central ones (via \eqref{eq:central-moments} and \eqref{eq:F-dec}),
\begin{equation}\label{eq:EGG-def-app}\begin{split}
   \EG_G(\alpha) 
   &= \sum_{\substack{a_1,..,a_N=0\\b_1,..,b_N=0}}^{\infty} \sum_{\substack{k_1,\ldots,k_N=0\\ l_1,\ldots,l_N=0}}^{\substack{a_1,\ldots,a_N\\ b_1\ldots,b_N}} F^{k_1,\ldots,k_N}_{l_1,\ldots,l_N} \prod^N_{j=1}\frac{(\alpha_{X_j})^{a_j}}{a_j!}\frac{(\alpha_{Y_j})^{b_j}}{b_j!} \binom{a_j}{k_j}\binom{b_j}{l_j} (-1)^{a_j-k_j+b_j-l_j} \braket{\hat{X}_j}^{a_j-k_j}\braket{\hat{Y}_j}^{b_j-l_j} 
   \\
   &= \sum_{\substack{a_1,..,a_N=0\\b_1,..,b_N=0}}^{\infty}  G^{a_1,..,a_N}_{b_1,...,b_N}\prod^N_{j=1}\frac{(\alpha_{X_j})^{a_j}}{a_j!}\frac{(\alpha_{Y_j})^{b_j}}{b_j!} \ ,
\end{split}\end{equation}
which we can then use as a definition alternative to \eqref{eq:EGG-predef-app}.
By $(i)$ rearranging the sums in the 1st line of \eqref{eq:EGG-def-app} via a generalization of the formula $\sum_{a=0}^{\infty}\sum_{k=0}^a A^{a,k} = \sum_{a=0}^{\infty}\sum_{k=0}^{\infty} A^{a+k,k}$ (for any abstract quantity $A^{a,k}$) to multi-indexes, $(ii)$ expanding the central moments via \eqref{eq:F-dec} and $(iii)$ splitting the powers of $\alpha_{X_j},\alpha_{Y_j}$ between the indexes $a_j,b_j,k_j,l_j$ we note, that said sum can be rewritten as a product of two series gathering up to two simple functionals
\begin{equation}\label{eq:EGG-gathered}
   \EG_G(\alpha) = \braket{e^{f_{\alpha}(\hat{X},\hat{Y}) }}e^{-\braket{f_{\alpha}(\hat{X},\hat{Y}) }} \ , 
   \qquad 
   f_{\alpha}(\hat{X},\hat{Y}) := \sum_{j=1}^N(\alpha_{X_j}\hat{X}_j+\alpha_{Y_J}\hat{Y}_j) \ .
\end{equation}

Having a generating function at our disposal, we can now proceed to compute Poisson brackets between moments. We do so in a way analogous to the derivation of the |OG{Poisson} brackets for absolute moments performed in the previous subsection. On the one hand, we implement to each argument a series expansion provided in the 2nd line of \eqref{eq:EGG-def-app}:
\begin{equation}\begin{split}\label{eqa:EGG-comm}
  \{ \EG_G(\alpha),\EG_G(\beta)\}=\sum_{\substack{a_1,..,a_N=0\\b_1,..,b_N=0\\c_1,..,c_N=0\\d_1,..,d_N=0}}^{\infty} \prod^N_{j=1}\frac{(\alpha_{X_j})^{a_j}}{a_j!}\frac{(\alpha_{Y_j})^{b_j}}{b_j!}\frac{(\beta_{X_j})^{c_j}}{c_j!}\frac{(\beta_{Y_j})^{d_j}}{d_j!} \{G^{a_1,..,a_N}_{b_1,...,b_N},G^{c_1,..,c_N}_{d_1,...,d_N}    \} \ .
\end{split}\end{equation}
On the other hand, we express the second argument in the form \eqref{eq:EGG-gathered} and apply the Leibniz's rule, which in turn yields
\begin{equation}\label{eq:EGG-comm2}
   \{ \EG_G(\alpha),\EG_G(\beta)\}= \{ \EG_G(\alpha),\braket{e^{f_{\beta}(\hat{X},\hat{Y})}}\}e^{-\braket{f_{\beta}(\hat{X},\hat{Y})}} + \{ \EG_G(\alpha),e^{-\braket{f_{\beta}(X,Y)}}\} \braket{e^{f_{\beta}(\hat{X},\hat{Y})}}\
\end{equation}

The righthand side of the above equation is a sum of two terms, among which the second one vanishes due to orthogonality of central moments and the expectation values of the observables they are built over. To show that orthogonality let us expand the central moments in the relevant Poisson bracket according to its definition \eqref{eq:central-moments}. This yields: 
\begin{equation}\label{eq:central-orthog}\begin{split}
 &\{ X_i,  G^{a_1,..,a_N}_{b_1,...,b_N} \}=\sum_{\substack{k_1=0,..,k_N=0\\m_1=0,..m_N=0}}^{\substack{a_1,...,a_N\\b_1,...,b_N}}\prod_{j=1}^NB_{k_j,m_j}^{a_j,b_j}X_j^{a_j-k_j} \{ X_i , Y_1^{b_1-m_1}...Y_N^{b_N-m_N}\braket{\hat{F_1}^{k,n}...\hat{F_N}^{k,n}} \}\\
 &=\sum_{\substack{k_1=0,..,k_N=0\\m_1=0,..m_N=0}}^{\substack{a_1,...,a_N\\b_1,...,b_N}}\prod_{j=1}^N B_{k_j,m_j}^{a_j,b_j}X_j^{a_j-k_j}\Big(  \{ X_i , Y_1^{b_1-m_1}...Y_N^{b_N-m_N}\}
\braket{\hat{F_1}^{k,n}...\hat{F_N}^{k,n}}+ Y_j^{b_j-m_j}\{ X_i , \braket{\hat{F_1}^{k,n}...\hat{F_N}^{k,n}} \}\  ,
\end{split}\end{equation}
where we again use Leibniz's rule to split the above expression into two independent terms. Investigating the first one results in:  
\begin{equation}\begin{split}\label{eq:first}
&\sum_{\substack{k_1=0,..,k_N=0\\m_1=0,..m_N=0}}^{\substack{a_1,...,a_N\\b_1,...,b_N}}\prod_{j=1}^N B_{k_j,m_j}^{a_j,b_j}X_j^{a_j-k_j}  \{ X_i , Y_1^{b_1-m_1}...Y_N^{b_N-m_N}\}
\braket{\hat{F_1}^{k,n}...\hat{F_N}^{k,n}}\\
&= \sum_{\substack{k_1=0,..,k_N=0\\m_1=0,..m_N=0}}^{\substack{a_1,...,a_N\\b_1,...,b_N}}\prod_{j=1}^N B_{k_j,m_j}^{a_j,b_j}X_j^{a_j-k_j} Y_1^{b_1-m_1}...\frac{A_i}{\hbar}(b_i-m_i)Y_i^{b_i-m_i-1}...Y_N^{b_N-m_N}
\braket{\hat{F_1}^{k,n}...\hat{F_N}^{k,n}}\\
&= \sum_{\substack{k_1=0,..,k_N=0\\m_1=0,..m_N=0}}^{\substack{a_1,...,a_N\\b_1,...,b_N}}\prod_{\substack{j=1\\ j\neq i}}^N B_{k_j,m_j}^{a_j,b_j}X_j^{a_j-k_j} Y_j^{b_j-m_j}\frac{A_i}{\hbar}b_i(-1)^{a_i+b_i-k_i-m_i}\binom{a_i}{k_i}\binom{b_i-1}{m_i}Y_i^{b_i-m_i-1}X_i^{a_i-k_i}
\braket{\hat{F_1}^{k,n}...\hat{F_N}^{k,n}}\\
&=-b_i\frac{A_i}{\hbar}G^{a_1,..,a_N}_{b_1,..,b_i-1,...,b_N} \ ,
\end{split}\end{equation}
where in the second line we use the Poisson bracket between $X_j$ and the function of $Y_j$. 
Manipulating the binomial coefficients in the third line, we further simplify the expression, ultimately representing it as a central moment lowered in the $b_i$ index in the last line.  

For the second term, we use the relation \eqref{eq:absolute-condition} defining the algebra structure of the absolute moments:
\begin{equation}\begin{split}
 &\sum_{\substack{k_1=0,..,k_N=0\\m_1=0,..m_N=0}}^{\substack{a_1,\ldots,a_N\\b_1,\ldots,b_N}}\prod_{j=1}^N B_{k_j,m_j}^{a_j,b_j}X_j^{a_j-k_j} Y_j^{b_j-m_j}\{ X_i , \braket{\hat{F_1}^{k,m}\ldots\hat{F_N}^{k,m}} \} = \\ 
 &=\sum_{\substack{k_1=0,\ldots,k_N=0\\m_1=0,\ldots,m_N=0}}^{\substack{a_1,\ldots,a_N\\b_1,\ldots,b_N}}\prod_{j=1}^N B_{k_j,m_j}^{a_j,b_j}X_j^{a_j-k_j}  Y_j^{b_j-m_j}\frac{1}{i\hbar}\braket{[ X_i , \hat{F_1}^{k,m}\ldots\hat{F_N}^{k,m}]} \\
 &=\sum_{\substack{k_1=0,\ldots,k_N=0\\m_1=0,\ldots,m_N=0}}^{\substack{a_1,\ldots,a_N\\b_1,\ldots,b_N}}\prod_{j=1}^N B_{k_j,m_j}^{a_j,b_j}X_j^{a_j-k_j} Y_j^{b_j-m_j}\frac{A_i}{\hbar}m_i\braket{\hat{F_1}^{k,m}\ldots\hat{F_i}^{k,m-1}\ldots\hat{F_N}^{k,m}} \\
 &=\sum_{\substack{k_1=0,\ldots,k_N=0\\m_1=0,\ldots,m_N=0}}^{\substack{a_1,\ldots,a_N\\b_1,\ldots,b_N}}\prod_{\substack{j=1\\ j\neq i}}^N B_{k_j,m_j}^{a_j,b_j}X_j^{a_j-k_j} Y_j^{b_j-m_j}\frac{A_i}{\hbar}(-1)^{a_i+b_i-k_i-m_i}\binom{a_i}{k_i}\binom{b_i}{m_i} \\
 &\hspace{3cm}\times m_i Y_i^{b_i-m_i}X_i^{a_i-k_i}\braket{\hat{F_1}^{k,m}\ldots\hat{F_i}^{k,m-1}\ldots\hat{F_N}^{k,m}} \ .
\end{split}\end{equation}
Upon reindexing the sum via a new set of indexes $m'_i=m_i-1$, the above expression reduces to the one obtained in \eqref{eq:first} but with the opposite sign. Thus, we conclude that the Poisson Bracket between $X_i$ variable and any central moment vanishes. The same reasoning can be applied to prove that the Poison bracket of $Y_j$ variable with any central moment vanishes as well. Thus
\begin{equation}
  \{ X_i,  G^{a_1,..,a_N}_{b_1,...,b_N} \}=\{ Y_i,  G^{a_1,..,a_N}_{b_1,...,b_N} \}=0 \ ,
\end{equation}
which implies in particular vanishing of the second term of the righthand side of \eqref{eq:EGG-comm2}. By plugging \eqref{eq:EGG-gathered} into the first term of that expression we get
\begin{equation}\label{eq:EGG-comm3}\begin{split}
   \{ \EG_G(\alpha),\EG_G(\beta)\} 
   &=\{e^{-\braket{f_{\alpha}(\hat{X},\hat{Y})}},\braket{e^{f_{\beta}(\hat{X},\hat{Y})}}\}\braket{e^{f_{\alpha}(\hat{X},\hat{Y})}}e^{-\braket{f_{\beta}(\hat{X},\hat{Y})}} \\ 
   &+ \{ \braket{e^{f_{\alpha}(\hat{X},\hat{Y})}},\braket{e^{f_{\beta}(\hat{X},\hat{Y})}}\}e^{-\braket{f_{\beta}(\hat{X},\hat{Y})}}e^{-\braket{f_{\alpha}(\hat{X},\hat{Y})}} \ .
\end{split}\end{equation}
The first Poisson bracket in the last line in the above formula acts between a function of classical variables and (an expectation value of the) generating function for the absolute moments \eqref{eq:F-gen} introduced in the previous subsection. By Taylor-expanding the first argument, implementing the middle formula of the first line of \eqref{eq:F-gen}, employing the formula \eqref{eq:X-F-comm} for the commutator between $(\hat{X}_j,\hat{Y}_j)$ and $(\hat{F}^{a,b})_j$ and gathering the series back, we obtain:
\begin{equation}\begin{split}\label{eq:first-p}
   \{e^{-\braket{f_{\alpha}(\hat{X},\hat{Y})}},\braket{e^{f_{\beta}(\hat{X},\hat{Y})}}\}=\sum_{j=1}^N \frac{A_j}{\hbar}(\beta_{X_j}\alpha_{Y_j}-\alpha_{X_j}\beta_{Y_j})e^{-\braket{f_{\alpha}(\hat{X},\hat{Y})}}\braket{e^{f_{\beta}(\hat{X},\hat{Y})}} \ .
\end{split}\end{equation}
Similarly, by expanding $\braket{e^{f_{\alpha}(\hat{X},\hat{Y})}}$ as the (expectation value of the) product of generating functions for absolute moments and using \eqref{eq:EGF-comm} we evaluate the second Poisson bracket: 
\begin{equation}\begin{split}\label{eq:second-p}
   \{ \braket{e^{f_{\alpha}(\hat{X},\hat{Y})}},\braket{e^{f_{\beta}(\hat{X},\hat{Y})}}\}=\frac{1}{i\hbar} \braket{ [e^{f_{\alpha}(\hat{X},\hat{Y})},e^{f_{\beta}(\hat{X},\hat{Y})}]}=\frac{2}{\hbar}\sin\Big(\sum_{j=1}^N \frac{A_j}{2}(\alpha_{X_j}\beta_{Y_j}-\alpha_{Y_j}\beta_{X_j})\Big)\braket{e^{f_{\alpha+\beta}(\hat{X},\hat{Y})}} \ .
\end{split}\end{equation}
Finally, by plugging \eqref{eq:first-p}, \eqref{eq:second-p} back into \eqref{eq:EGG-comm3} and simplification we get an alternative to \eqref{eqa:EGG-comm} expression for the Poisson bracket between generating functions for central moments
\begin{equation}\begin{split}
   \{ \EG_G(\alpha),\EG_G(\beta)\} &= \sum_{j=1}^N \frac{A_j}{\hbar}(\beta_{X_j}\alpha_{Y_j}-\alpha_{X_j}\beta_{Y_j})\EG_G(\alpha)\EG_G(\beta) \\
   &+ \frac{2}{\hbar}\sin\Big(\sum_{j=1}^N \frac{A_j}{2}(\alpha_{X_j}\beta_{Y_j}-\alpha_{Y_j}\beta_{X_j})\Big)\EG_G(\alpha+\beta) 
   = : \PBG_2 + \PBG_1 \ ,
\end{split}\end{equation}
where $\PBG_1$ and $\PBG_2$ are the terms linear and quadratic in $\EG_G$ (second and first term of the middle expression) respectively.

In order to compare the above form with \eqref{eqa:EGG-comm}, we need to Taylor-expand generating functions present in $\PBG_1,\PBG_2$. Let us start with the second one, where after expanding we further rearrange the order of summation 
\begin{equation}
    \begin{split}\label{eq:EGG-p1}
        \PBG_2 &= 
        \sum_{n=1}^N \frac{A_n}{\hbar}\sum_{\substack{a_1,..,a_N=0\\b_1,..,b_N=0\\c_1,..,c_N=0\\d_1,..,d_N=0}}^{\infty}\prod^N_{\substack{j=1\\j\neq n}}  \frac{(\alpha_{X_j})^{a_j}}{a_j!}\frac{(\alpha_{Y_j})^{b_j}}{b_j!}\frac{(\beta_{X_j})^{c_j}}{c_j!}\frac{(\beta_{Y_j})^{d_j}}{d_j!} \\
        &\hspace{2cm}\times \Big(\frac{(\alpha_{X_n})^{a_n}}{a_n!}\frac{(\alpha_{Y_n})^{b_n+1}}{b_n!}\frac{(\beta_{X_n})^{c_n+1}}{c_n!}\frac{(\beta_{Y_n})^{d_n}}{d_n!}-\frac{(\alpha_{X_n})^{a_n+1}}{a_n!}\frac{(\alpha_{Y_n})^{b_n}}{b_n!}\frac{(\beta_{X_n})^{c_n}}{c_n!}\frac{(\beta_{Y_n})^{d_n+1}}{d_n!}\Big) \\ 
        &\hspace{2cm}\times G^{a_1,..,a_N}_{b_1,...,b_N} G^{c_1,..,c_N}_{d_1,...,d_N}\\
        &=\sum_{\substack{a_1,..,a_N=0\\b_1,..,b_N=0\\c_1,..,c_N=0\\d_1,..,d_N=0}}^{\infty}\prod^N_{\substack{j=1\\j\neq n}}\frac{(\alpha_{X_j})^{a_j}}{a_j!}\frac{(\alpha_{Y_j})^{b_j}}{b_j!}\frac{(\beta_{X_j})^{c_j}}{c_j!}\frac{(\beta_{Y_j})^{d_j}}{d_j!}\\
        &\hspace{2cm}\times \sum_{n=1}^N \frac{A_n}{\hbar}\Big(b_n c_nG^{a_1,..,a_N}_{b_1,...,b_n-1,...,b_N}G^{c_1,..,c_n-1,...,c_N}_{d_1,...,d_N}-a_n d_nG^{a_1,..,a_n-1,...,a_N}_{b_1,...,b_N}G^{c_1,..,c_N}_{d_1,...,d_n-1,...,d_N}\Big) \ .
    \end{split}
\end{equation}
Similarly, for the first term $PB_1$ we Taylor-expand both the generating function and the sine function, further decomposing the powers of sums by binomial theorem 
\begin{equation}\begin{split}
\PBG_1 &= 
\frac{2}{\hbar}\sum_{r=0}^{\infty}\frac{(-1)^r}{(2r+1)!}\Big(\sum_{j=1}^N \frac{A_j}{2}(\alpha_{X_j}\beta_{Y_j}-\alpha_{Y_j}\beta_{X_j})\Big)^{2r+1}\sum_{\substack{m_1,..,m_N=0\\n_1,..,n_N=0}}^{\infty}\prod^N_{\substack{j=1}} \frac{(\alpha_{X_j}+\beta_{X_j})^{m_j}}{m_j!}\frac{(\alpha_{Y_j}+\beta_{Y_j})^{n_j}}{n_j!}G^{n_1,...,n_N}_{m_1,...,m_N}\\
&=\frac{1}{\hbar}\sum_{r=0}^{\infty}\sum_{s=0}^{2r+1}\frac{(-1)^{r+s}}{4^r}\sum_{k_1=0,..,k_N=0}^{2r+1-s}\delta_{2r+1-s,k_1+..+k_N}\sum_{l_1=0,...,l_N=0}^s\delta_{s,l_1+..+l_N}\\
&\hspace{1cm}\times \sum_{\substack{a_j=0, d_j=0\\b_j=0,c_j=0}}^{\infty}\prod_{j=1}^N A_j^{k_j+l_j}\frac{(\alpha_{X_j})^{a_j+k_j}}{a_j!}\frac{(\alpha_{Y_j})^{b_j+l_j}}{b_j!}\frac{(\beta_{X_j})^{c_j+l_j}}{c_j!}\frac{(\beta_{Y_j})^{d_j+k_j}}{d_j!}G^{a_1+c_1,...,a_N+c_N}_{b_1+d_1,...,b_N+d_N} \ .
\end{split}\end{equation}
Now, in order to bring the above expression to the form most similar to that of \ref{eqa:EGG-comm} we introduce new indexes: $a'_j=a_j+k_j$, $d'_j=d_j+k_j$, $c'_j=c_j+l_j$, $b'_j=b_j+l_j$ and change the order of summation so that the summations over all $a'$, $b'$, $c'$ and $d'$ are brought to the front. The result yields
\begin{equation}\label{eq:PBG-tmp1}\begin{split}
\PBG_1 
&=\frac{1}{\hbar}\sum_{\substack{a'_1,..,a'_N=0\\b'_1,..,b'_N=0\\c'_1,..,c'_N=0\\d'_1,..,d'_N=0}}^{\infty} \prod^N_{j=1}\frac{(\alpha_{X_j})^{a'_j}}{a'_j!}\frac{(\alpha_{Y_j})^{b'_j}}{b'_j!}\frac{(\beta_{X_j})^{c'_j}}{c'_j!}\frac{(\beta_{Y_j})^{d'_j}}{d'_j!}\sum_{r=0}^{\infty}\sum_{s=0}^{2r+1}\frac{(-1)^{r+s}}{4^r}\\
& \times  \sum_{k_1,..,k_N}\delta_{2r+1-s,k_1+..+k_N}\sum_{l_1,...,l_N}\delta_{s,l_1+..+l_N} k_j ! l_i !  \prod^N_{j=1}A_j^{k_j+m_j} \binom{a'_j}{k_j}\binom{b'_j}{l_j}\binom{c'_j}{l_j}\binom{d'_j}{k_j}G^{a'_1+c'_1-k_1-l_1,...,a'_N+c'_N-k_N-l_N}_{b'_1+d'_1-k'_1-l_1,...,b'_N+d'_N-k_N-l_N} \ ,
\end{split}\end{equation}
with indices  $k_i$ and $l_i$ running over the sets: 
\begin{equation}
    0 \leq k_i \leq \min(2r+1-s,a_i,d_i) \ , \qquad 0 \leq l_i \leq \min(s,b_i,c_i) \ .
\end{equation}
By introducing a new index $q_i=k_i+l_i$ and the following function of coefficients: 
\begin{equation}\label{eq:K}
    K^{a_i,b_i,c_i,d_i}_{q_i,s}=\sum_{l_1,...,l_N}\delta_{s,l_1+..+l_N}  \prod^N_{i=1}l_i!(q_i-l_i)!\binom{a_i}{q_i-l_i}\binom{b_i}{l_i}\binom{c_i}{l_i}\binom{d_i}{q_i-l_i} \ ,
\end{equation}
the last form of $\PBG_1$ in \eqref{eq:PBG-tmp1} can be brought to the slightly simpler form
\begin{equation}\label{eq:PB1-last}
\begin{split}
\PBG_1 &= 
\frac{1}{\hbar}\sum_{\substack{a_1,..,a_N=0\\b_1,..,b_N=0\\c_1,..,c_N=0\\d_1,..,d_N=0}}^{\infty} \prod^N_{j=1}\frac{(\alpha_{X_j})^{a_j}}{a_j!}\frac{(\alpha_{Y_j})^{b_j}}{b_j!}\frac{(\beta_{X_j})^{c_j}}{c_j!}\frac{(\beta_{Y_j})^{d_j}}{d'_j!}\sum_{r=0}^{\infty}\sum_{s=0}^{2r+1}\frac{(-1)^{r+s}}{4^r}  \\ 
&\hspace{3cm}\times \sum_{q_1,\ldots,q_N}\delta_{2r+1,q_1+..+q_N}A_j^{q_j} K^{a_i,b_i,c_i,d_i}_{q_i,s} G^{a_1+c_1-q_1,\ldots,a_N+c_N-q_N}_{b_1+d_1-q_1,\ldots,b_N+d_N-q_N} \ ,
\end{split}
\end{equation}
with indices now running over the sets
\begin{subequations}\begin{align}
   \max(0,q_i-\min(2r+1-s,a_i,d_i)) &\leq l_i \leq \min(q_i,\min(s,b_i,c_i)) \ , 
   \\
   0 &\leq q_i \leq \min(s,b_i,c_i)+\min(2r+1-s,a_i,d_i) \ .
\end{align}\end{subequations}
Finally, comparing \eqref{eq:EGG-p1} and \eqref{eq:PB1-last} with \ref{eqa:EGG-comm} order by order we identify the closed formula for Poisson brackets between two central moments:
\begin{equation}\begin{split}
  \{G^{a_1,\ldots,a_N}_{b_1,\ldots,b_N}, G^{c_1,\ldots,c_N}_{d_1,\ldots,d_N} \}
  &= \sum_{n=1}^N \frac{A_n}{\hbar}\Big(b_n c_nG^{a_1,\ldots,a_N}_{b_1,\ldots,b_n-1,\ldots,b_N}G^{c_1,\ldots,c_n-1,\ldots,c_N}_{d_1,\ldots,d_N}-a_n d_nG^{a_1,\ldots,a_n-1,\ldots,a_N}_{b_1,\ldots,b_N}G^{c_1,\ldots,c_N}_{d_1,\ldots,d_n-1,\ldots,d_N}\Big)\\
  &+\frac{1}{\hbar}\sum_{r}\sum_{s}\frac{(-1)^{r+s}}{4^r}  \sum_{q_1,\ldots,q_N}\delta_{2r+1,q_1+\ldots+q_N}A_j^{q_j} K^{a_i,b_i,c_i,d_i}_{q_i,s} G^{a_1+c_1-q_1,\ldots,a_N+c_N-q_N}_{b_1+d_1-q_1,\ldots,b_N+d_N-q_N} \ ,
\end{split}
\end{equation}
with $K^{a_i,b_i,c_i,d_i}_{q_i,s}$ given by formula \eqref{eq:K} and (integer) indices running over the sets:
\begin{subequations}\begin{align}
    &\max(0,q_i-\min(2r+1-s,a_i,d_i)) \leq l_i \leq \min(q_i,\min(s,b_i,c_i))\\
   & 0 \leq q_i \leq \min(s,b_i,c_i)+\min(2r+1-s,a_i,d_i)\\
   & 0 \leq s \leq \min(2r+1,\sum_{i=1}^N \min(b_i,c_i))\\
   & 0 \leq r \leq \frac{1}{2}\sum_{i=1}^N \min(b_i,c_i)+\frac{1}{2}\sum_{i=1}^N \min(a_i,d_i)-\frac{1}{2}
\end{align}\end{subequations}

\bibliography{bibliography}
\end{document}